\documentclass[12pt]{iopart} 
\usepackage{graphicx}

\newcommand{\be}{\begin{equation}}
\newcommand{\ee}{\end{equation}}
\newcommand{\ba}{\begin{eqnarray}}
\newcommand{\ea}{\end{eqnarray}}
\newcommand{\Mc}{{\cal M}}
\newcommand{\Ms}{M_{\odot}}
\newcommand{\m}{\langle}
\newcommand{\M}{\rangle}

\newcommand{\prd}{PRD }

\newcommand{\apj}{ApJ }
\newcommand{\mnras}{MNRAS }
\newcommand{\aap}{AAP}
\newcommand{\apss}{APSS}
\newcommand{\pasp}{PASP}
\newcommand{\bml}{\begin{mathletters}}
\newcommand{\eml}{\end{mathletters}}

\def\ltsima{$\; \buildrel < \over \sim \;$}
\def\simlt{\lower.5ex\hbox{\ltsima}}
\def\gtsima{$\; \buildrel > \over \sim \;$}
\def\simgt{\lower.5ex\hbox{\gtsima}}

\begin{document}

\title[The LISA verification binaries]{The LISA verification binaries}

\author{A. Stroeer and A. Vecchio}

\address{School of Physics and Astronomy, 
University of Birmingham, Edgbaston, Birmingham B15 2TT, UK
}
\ead{astroeer@star.sr.bham.ac.uk, av@star.sr.bham.ac.uk}
\begin{abstract}
The Laser Interferometer Space Antenna (LISA) guarantees the detection of gravitational waves by monitoring a handful of known nearby galactic binary systems, the so-called ``verification binaries''. We consider the most updated information on the source parameters for the thirty more promising verification binaries. We investigate which of them are indeed guaranteed sources for LISA and estimate the accuracy of the additional information that can be extracted during the mission. Our analysis considers the two independent Michelson outputs that can be synthesised from the LISA constellation, and we model the LISA transfer function using the rigid adiabatic approximation. We carry out extensive Monte Carlo simulations to explore the dependency of our results on unknown or poorly constrained source parameters. We find that four sources -- RXJ0806.3+1527, V407 Vul, ES Cet and AM CVn -- are clearly detectable in one year of observation; RXJ0806.3+1527 should actually be observable in less than a week. For these sources LISA will also provide information on yet unknown parameters with an error between $\approx$ 1\%  and  $\approx$ 10\%. Four additional binary systems -- HP Lib, 4U 1820-30, WZ Sge and KPD 1930+2752 -- might also be marginally detectable.
\end{abstract}


\section{Introduction}
\label{sec:int}

The Laser Interferometer Space Antenna (LISA) is an all-sky gravitational wave (GW) observatory in the frequency window $10^{-4} - 1$ Hz with launch date in the 2015+ time frame. LISA is expected to observe gravitational radiation from a variety of sources~\cite{Bender1998a,Cutler2002a}, such as stellar mass binary systems and black hole binaries over a large mass spectrum. LISA guarantees the detection of GW's by monitoring several stellar mass (primarily white dwarfs) binary systems  that are well known from electromagnetic observations and whose radiation is estimated to be sufficiently strong to be detected. They have been titled ``verification binaries'' and comprise systems from several classes: (i) mass-transferring AM CVn binary systems -- consisting of a low mass donor star and a higher mass white dwarf accretor -- in which mass transfer drives the orbital evolution to longer periods; (ii) double white dwarf binary systems, whose orbital evolution is primarily driven by gravitational radiation reaction, but possibly also tidal effects; (iii) (ultra-)compact X-ray binaries, in which a neutron star accretes material from a low mass orbital companion; and (iv) cataclysmic variables (CV), consisting of an accreting white dwarf and (typically) a low-mass main-sequence star. 

Verification binaries are crucial because they guarantee the direct detection of GW's and provide a new opportunity to study the complex physics of compact objects. Previous analyses~\cite{Phin2001a} of the detectability of verification binaries relied on crude estimates of their source parameters that are becoming progressively out of date due to new observations. As the mission formulation phase progresses, and firmer plans for data analysis are put in place, a more detailed study is needed to identify those systems that can be indeed considered ``verification binaries'' ({\em i.e.} guaranteed sources) and to explore what kind of new information LISA will be able to provide. 

In this paper we consider the most updated information on the thirty most promising nearby galactic binaries; we carry out a systematic study of the signal-to-noise ratio (SNR) at which they can be observed and explore the accuracy with which LISA can measure the unknown source parameters, keeping into account prior information. Our analysis shows that four systems (RXJ0806.3+1527, V407 Vul, ES Cet and AM CVn) can indeed be considered guaranteed sources and, as such, verification binaries. Four others (HP Lib, 4U 1820-30, WZ Sge and KPD 1930+2752) might also be marginally detectable; this will depend on the actual polarisation of gravitational waves, the duration of the mission, the exact noise level and the performance of data analysis algorithms. We also show that for RXJ0806.3+1527, V407 Vul, ES Cet and AM CVn, LISA will provide new information on unknown parameters (such as the orbital inclination angle and a combination of the distance and chirp mass) with an error $\approx$ 1\% - 10\%; this will be important to shed new light on the physics at work in such compact objects. The remaining binaries appear to be beyond the reach of LISA.

\section{Sources and signal model}

A large number of nearby galactic compact binary systems have been discovered and monitored by recent surveys; 30 of them have been identified as promising (or certain) candidates for GW detection with LISA: we will refer to them as {\em potential} verification binaries. In this section we introduce the sources and the relevant information that are currently available. We also review the model of the GW signal that is used in the next section to study the expected SNR characterising LISA observations and the estimates of the statistical errors associated to the measurement of unknown parameters.

\begin{table}
\begin{center}
\scriptsize{
\begin{tabular}{|l|cc|c|c|c|c|}
\hline
                        &\multicolumn{2}{|c|}{{\bf{Coordinates } }}  &{\bf{$f$}} [mHz]   &{\bf{$D$ [pc] }}    &  {\bf{$m_2$ [$\Ms$] }}   &     {\bf{$m_1$ [$\Ms$]}}\\
\hline
\multicolumn{7}{|c|}{{\bf{AM CVn systems}}}\\
\hline
{\bf{RX J0806.3+1527}}   &      120.443557 &-4.704035    & 6.2202766  &   300-1000  &       0.13     &        0.2-0.5$^{\cite{Isr2002a}}$ \\
{\bf{V407 Vul }}         &      294.994600 &+46.783096   & 3.51250    &   300-1000  &       0.068    &        0.7$^{\cite{Crop1998a}}$\\
{\bf{ES Cet     }}       &      25.310039  &-20.333218   & 3.22       &   350-1000  &       0.062    &        0.7$^{\cite{Esp2005a}}$\\
{\bf{AM CVn       }}     &      171.084432 &+37.441925   & 1.94414    &   606       &       0.14     &        0.85\\
{\bf{HP Lib         }}   &      235.786580 &+4.959494    & 1.813      &   197       &       0.032    &        0.57\\
{\bf{CR Boo      }}      &      202.971814 &+17.896456   & 1.360      &   337       &       0.023    &        0.55\\
{\bf{KL Dra       }}     &      334.830867 &+78.322322   & 1.333      &   100$^{\star}$       &       0.022    &        0.27\\
{\bf{V803 Cen      }}    &      216.866251 &-30.317527   & 1.241      &   347       &       0.021    &        1.31\\
{\bf{SDSS J0926+3624 }}    &      132.286781 &+20.234177   & 1.177      &   100$^{\star}$       &       0.02$^{\star}$     &        0.6$^{\star}$\\
{\bf{CP Eri        }}    &      42.830979  &-26.426918   & 1.176      &   100$^{\star}$       &       0.019    &        0.63\\
{\bf{2003aw        }}    &      140.733734 &-21.234213   & 0.9862     &   100$^{\star}$       &       0.015$^{\star}$     &        0.42$^{\star}$\\
{\bf{SDSS J1240-0159 }}   &      190.193388 &+2.225887    & 0.8921     &   350-440   &       0.015$^{\star}$     &        0.38$^{\star}$\\
{\bf{GP Com       }}     &      188.418156 &+23.000197   & 0.7158     &   75        &       0.010    &        0.45\\
{\bf{CE 315       }}     &      206.451728 &-14.462612   & 0.5120     &   77        &       0.006    &        0.48\\
\hline
\multicolumn{7}{|c|}{{\bf{(Ultra-)compact X-ray binaries}}}\\
\hline
{\bf{4U 1820-30 }}       &      275.841980 &-7.027394$^{\cite{chan2005a}}$  &  2.92$^{\cite{Kell1989a}}$   &      8100$^{\cite{Phin2001a}}$ &           $<$0.1$^{\cite{Phin2001a}}$      &      1.4$^{\cite{Phin2001a}}$\\
{\bf{4U 1543-624}}& 251.159958& -41.352944 & 1.832 &    5000$^{\star}$ &   0.04 & 1.4 \\
{\bf{4U 1850-087}}& 283.530386 &+14.112497 & 1.618 &    8200 &   0.03 & 1.4 \\
{\bf{4U 1626-67 }}       &      259.038082 &-44.908517 &  0.4$^{\cite{Wern2003a}}$    &      8000$^{\cite{Phin2001a}}$ &           0.02-0.08$^{\cite{Wern2003a}}$ &      1.4$^{\cite{Wern2003a}}$\\
{\bf{CC Com    }}        &      174.038625 &+21.775780$^{\cite{chan2005a}}$ &  0.105  &      90$^{\cite{Phin2001a}}$   &           0.36$^{\cite{Zhou1988a}}$      &      0.62$^{\cite{Zhou1988a}}$\\
\hline
\multicolumn{7}{|c|}{{\bf{Double white dwarfs}}}\\
\hline
{\bf{WD 0957-666 }}     &       209.226640 &-67.301935  & 0.379520267 & 135     &        0.32       &     0.37\\
{\bf{KPD0422+4521  }}   &       76.322155  &+22.854567$^{\cite{chan2005a}}$  & 0.256690$^{\cite{Koen1998a}}$    & 100$^{\star}$     &        0.511$^{\cite{Osz1999a}}$      &     0.526$^{\cite{Osz1999a}}$\\
{\bf{KPD1930+2752  }}   &       302.454582 &+48.904889$^{\cite{chan2005a}}$  & 0.2434262$^{\cite{Wool2002a}}$   & 100$^{\star}$     &        0.5$^{\cite{Wool2002a}}$        &     0.97$^{\cite{Wool2002a}}$\\
{\bf{WD 1101+364   }}   &       152.950020 &+27.689153  & 0.15996     & 97      &        0.31$^{\cite{Marsh1995a}}$       &     0.36$^{\cite{Marsh1995a}}$\\
{\bf{WD 1704+481  }}    &       242.213221 &+70.224728$^{\cite{chan2005a}}$  & 0.1598789$^{\cite{Max1999a}}$   & 100$^{\star}$     &        0.39$^{\cite{Max1999a}}$       &     0.56$^{\cite{Max1999a}}$\\
{\bf{WD 2331+290   }}   &       7.631717   &+29.202332 $^{\cite{chan2005a}}$ & 0.1 $^{\cite{Marsh1995a}}$        & 100$^{\star}$     &        $>$0.32 $^{\cite{Marsh1995a}}$     &     0.39$^{\cite{Marsh1995a}}$\\
\hline
\multicolumn{7}{|c|}{{\bf{Cataclysmic variables}}}\\
\hline
{\bf{EI Psc}}& 356.370891&+8.925019 &   0.5195&      210 &    0.13& 0.7 \\
{\bf{SDSS J1507+5230}} &192.423733 &+64.768925  & 0.4958 &  100$^{\star}$  &   0.13$^{\star}$ &   0.7$^{\star}$\\
{\bf{GW Lib}}&234.794265 & -6.417916 &  0.4341 &      100$^{\star}$ & 0.13$^{\star}$ &  0.7$^{\star}$ \\
{\bf{WZ Sge   }}        & 309.727920 &+36.921088 &  0.4065 &      43   &           $<$0.11   &        $>$0.7\\
{\bf{SDSS J0903+3300}} &128.734149 &+15.542558  & 0.3918 & 100$^{\star}$  &   0.13$^{\star}$ &  0.7$^{\star}$ \\
\hline
\end{tabular}
}
\end{center}
\caption{The 30 close stellar mass binary systems that are currently regarded as potential ``verification binaries'' for LISA. The table summarises the relevant parameters for LISA observations: ecliptic coordinates (J2000.0), gravitational wave frequency $f$, distance $D$ to the source and masses $m_1$ and $m_2$ of the two stars. The table is based on the observational summary reported in~\cite{VerBin2006a} and provides explicit references where additional information have been included. For those systems whose distance is not known we set $D = 100$ pc, a {\em lower limit} to $D$ due to the lack of detection of proper motion. For 4U 1543-624 we adopt $ D = 5000$~pc~\cite{VerBin2006a} due to its astrophysical nature. These values are marked by a superscript star. In six cases --  SDSS J0926, 2003aw, SDSS J1240, SDDS J1507+5230, GW Lib and SDDS J0903+3300 -- the mass determination is highly uncertain or not available. For them (again marked by a superscript star) we adopt canonical values for this class of systems and keep into account estimates of the mass ratio (see the text for more details). In the table we classify RX J0806.3+1527 and V407 Vul as AM CVn systems, but we note that their classification is still debated, see~\cite{VerBin2006a} and references therein.}
\label{tab:dataset}
\end{table}

In Table~\ref{tab:dataset} we summarise the 30 most promising sources for LISA and their parameters: more information, including references can be found in~\cite{VerBin2006a}. The location of the systems in the sky -- identified by the unit vector $\vec{N}$ and ecliptic coordinates $(\theta_{N},\varphi_{N})$ -- and the orbital period (and therefore the frequency $f$ of GWs%
\footnote{GW's are emitted at all the multiples of the orbital frequency: here we just consider the leading order Newtonian mass quadrupole radiation, that provides by far the dominant contribution.})
are extremely well determined; they can be considered (at least for the analysis presented in this paper) exactly known. On the other hand, some parameters that are essential to evaluate the expected SNR are either unknown or poorly constrained. For none of the sources we have information about the orientation of the binary orbital angular momentum $\vec{L}$ (described by the polar coordinates $\theta_L$ and $\varphi_L$), which determines the signal polarisation, and in turn the actual GW amplitude at the detector through the antenna beam patterns. For several systems we also have only poor (if any) information about the distance to the source $D$ and/or the masses of the stars $m_1$ and $m_2$, which affect the GW amplitude. In this paper, we adopt a distance $D = 100$ pc  for KL Dra, SDSS J0926, CP Eri, 2003aw, KPD 0422+4521, KPD 1930+2752, WD 1704+481, WD 2331+290,SDDS J1507+5230, GW Lib and SDDS J0903+3300; note that this value represents a {\em lower limit} to $D$, because the proper motion of those objects is not observed. For 4U 1543-624 we adopt a distance $D=5$ kpc \cite{VerBin2006a}, based on evolutionary arguments. For SDDS J0926 no information on the masses are available: in our analysis we assume typical mass values for AM CVn binaries~\cite{Marsh2004a}, {\em i.e.} $m_2 = 0.02\,\Ms$ and $m_1 =  0.6\,\Ms$. For 2003aw and SDDS J1240, some measurements of the mass ratio have been obtained ($m_2/m_1\approx 0.036$ and $\approx 0.039$ for 2003aw and SDSS J1240, respectively): by keeping into account these constraints and using a standard AM CVn model~\cite{Marsh2004a} we adopt the mass values as reported in Table~\ref{tab:dataset}. For the three CV systems SDDS J1507+5230, GW Lib and SDDS J0903+3300, whose masses have not been measured, we set the values of $m_1$ and $m_2$ to those of EI Psc, the only CV with a reliable mass determination.

In our analysis we consider the output of the two uncorrelated Michelson channels as introduced in~\cite{Cutler1998a} and include the effect of the LISA transfer function using the rigid adiabatic approximation~\cite{Rubbo2004a}. The gravitational wave signal $h(t; \vec{\lambda})$ at each of the two readouts can therefore be written as~\cite{Vecc2004a}
\be
h(t; \vec{\lambda})=\sum_{n=1}^4 B_n(t) \cos\chi_n(t)\,,\nonumber\\
\label{e:ht}
\ee
where
\ba
B_n(t)\! & = &\! \!\left[(F_{n}^{(+)}(t)\,A_{+}(t))^{2} + (F_{n}^{(\times)}(t)\,A_{\times}(t))^{2}\right]^{1/2}\,,
\label{B}
\\
\chi_n(t) &\! =& \!\!\phi_{\rm gw}(t) + \phi_{n}(t) + \phi_{{\rm D}}(t)\,.
\label{chi}
\ea
The Doppler phase shift $\phi_{{\rm D}}(t)$ is caused by the motion of the interferometer around the sun and the terms 
\be
\phi_{n}(t) = \arctan\left[-\frac{F_{n}^{(\times)}(t)\,A_{\times}(t)}
{F_{n}^{(+)}ç(t)\,A_{+}(t)}\right]
\label{polphase}	
\ee
are the polarisation phases induced by the change of LISA's orientation. $F_{n}^{(+,\times)}(t)$ are the antenna beam patters. Note that $\phi_n(t)$ and $F_{n}^{(+,\times)}(t)$ are different for the two Michelson channels. The amplitudes of the two independent polarisations in Eq.~(\ref{B}) are given by 
\ba
A_{+}(t) & = & 2\, \frac{\Mc^{5/3}}{D}\,
\left[ 1 + \left({\vec{L}} \cdot {\vec{N}}\right)^2\right]\,
\left[\pi\,f(t)\right]^{2/3}\,,
\label{e:Aplus}
\\
A_{{\times}}(t) & = & - 4\, \frac{\Mc^{5/3}}{D}\,
\left({\vec{L}} \cdot {\vec{N}}\right)\,
\left[\pi\,f(t)\right]^{2/3}\,,
\label{e:Across}
\ea
where the chirp mass is defined as $\Mc \equiv (m_1 + m_2)^{2/5}\,[(m_1 \, m_2)/(m_1 + m_2)]^{3/5}$. The amplitudes depend on the inclination angle  $\iota$ given by
\be
\cos \iota= {\vec{L}} \cdot {\vec{N}} 
=\cos\theta_L \cos\theta_{N} 
+ \sin\theta_{L}\sin\theta_{N} \cos(\varphi_{L}-\varphi_{N})\,.
\label{e:iota}
\ee

In this paper we model the gravitational wave signal from the sources in Table~\ref{tab:dataset} as exactly monochromatic with respect to an observer at rest in the solar system barycentre. In reality the frequency does change due to radiation reaction, mass transfer and tidal effects. However, we estimate that for the vast majority of the sources considered here the frequency drift will not be observable during the LISA mission. Exceptions are likely to be RXJ0806.3+1527 and V407 Vul, where measurements of the period derivatives have already been reported~\cite{VerBin2006a}; in fact, LISA could provide new direct measurements of their orbital evolution and details on the physics at work. Neglecting this intrinsic frequency drift (as we do in the present analysis) does not introduce any sizable effect on the SNR that is computed in the next section, and it affects the estimates of the statistical errors of the measurements only in a negligible way compared to the present uncertainties on the source parameters, see~\cite{Stroeer2006a}. In summary the phase of the gravitational wave signal in Eq.~(\ref{chi}) is given by
\be
\phi_{\rm gw}(t) = 2 \pi f_0 t + \Phi_0\,,
\label{e:phigw}
\ee
where $f_0$ and $\Phi_0$ are the frequency and phase, respectively, at the beginning of the observations. It is also useful to introduce the constant amplitude 
\be
A = 2\,\left(\pi\,f_0\right)^{2/3} \frac{\Mc^{5/3}}{D}\,.
\label{e:A}
\ee
The signal $h(t; \vec{\lambda})$ that we consider is hence described by the parameter vector
\be
\vec{\lambda}=\{\ln A,\Phi_{0},f_0,\cos \theta_{N},\varphi_{N},\cos \theta_{L},\varphi_{L}\}\,.
\ee

We introduce the usual inner product between two signals $h$ and $g$ as~\cite{Cutler1998a}:
\be
\left(h|g\right) =  2\,\int_{-\infty}^{+\infty} 
\frac{\tilde h^*(f) \tilde g(f)}{S_n(f)}\,df
\simeq \frac{2}{S_0}\,\int_{-\infty}^{+\infty} 
h(t) g(t)\,dt\,,
\label{inner}
\ee
where the latter equality follows from Parseval's theorem; $S_n(f)$ is the one-sided noise power spectral density that we consider constant around $f_0$; we introduce the notation $S_0 = S_n(f_0)$. In our analysis we model the noise spectral density according to the latest estimate of the instrumental and ``confusion noise''~\cite{Barack2004a}. In the noise produced by astrophysical foregrounds we include unresolved galactic and extra-galactic white dwarf binaries, but not the contribution from captures of solar-mass compact objects by massive black holes. If such foreground radiation is indeed present at the level described in~\cite{Barack2004a}, the SNR at which LISA can detect the verification binaries with $f_0$ in the mHz region would be reduced by a factor $\simlt 1.5$ with respect to what we report in the next section. The optimal signal-to-noise ratio $\rho$ at which $h(t; \vec{\lambda})$ can be detected is given by:
\be
\rho^2 = \sum_{a = 1}^2 (h^{(a)}|h^{(a)})\,,
\label{snr}
\ee
where $a = 1,2$ labels the Michelson outputs. The statistical error  $\Delta \vec{\lambda}$ associated to the measurement of $\vec{\lambda}$ can be computed using the variance-covariance matrix. In the limit $\rho \gg 1$, $\Delta \vec{\lambda}$ follows the Gaussian probability distribution
\be
p(\Delta \vec{\lambda}) = \left(\frac{\textrm{det}({\bf{\Gamma}})}{2 \,\pi}\right)^{1/2}
\,e^{-\frac{1}{2}\,\Gamma_{jk} \Delta\lambda^j \Delta\lambda^k}, 
\ee
where the Fisher information matrix is given by
\be
 \Gamma_{jk} \equiv \sum_a
\left(\frac{\partial h^{(a)}}{\partial \lambda^j} \Biggl|\Biggr.
\frac{\partial h^{(a)}}{\partial \lambda^k}\right)\,.
\label{fisher}
\ee
The (lower limit to the) expected mean square error on each parameter can be computed as 
\be
\m (\Delta\lambda^j)^2\M = \left[\left({\bf \Gamma}\right)^{-1}\right]^{jj}\,.
\label{mse}
\ee
Note that the errors depend on the actual value of $\vec{\lambda}$ and scale with the SNR according to $\m (\Delta\lambda^j)^2\M \propto 1/\rho^2$.

\begin{figure}
\vspace{0.3cm}
\resizebox{7.5cm}{!}{\includegraphics{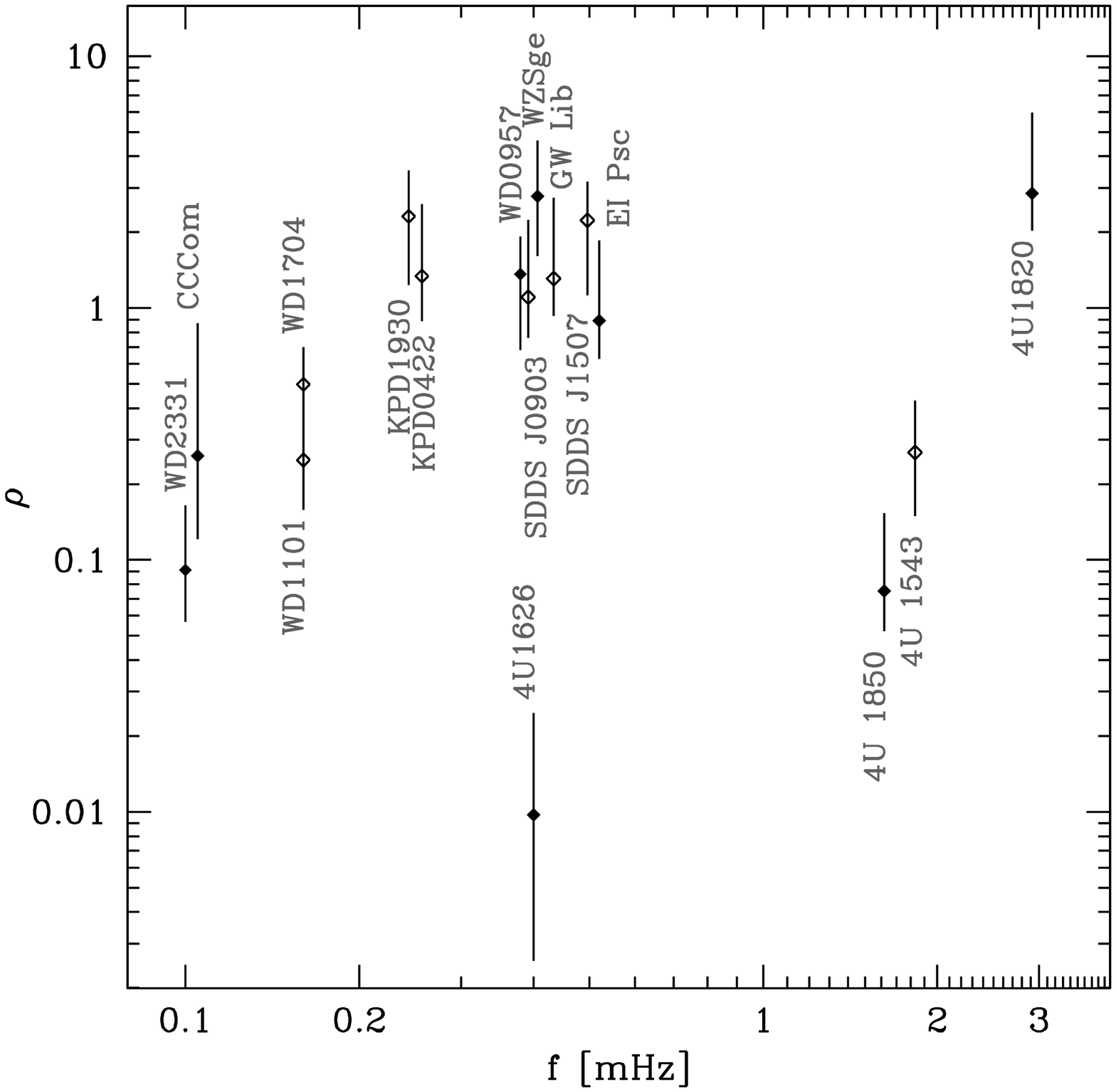}}
\resizebox{7.5cm}{!}{\includegraphics{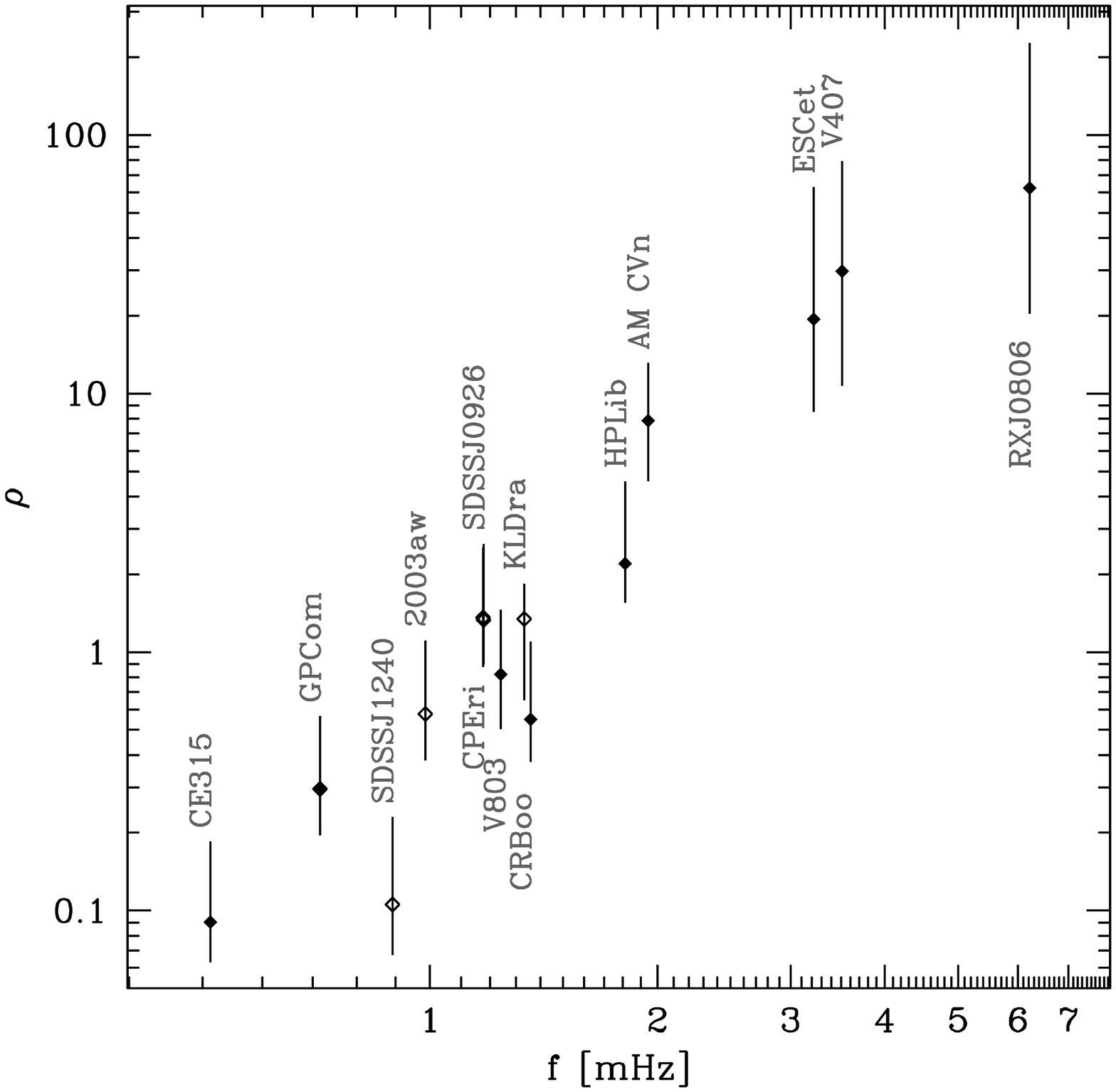}}
\caption{The signal-to-noise ratio at which LISA can observe the known binaries in one year of observation. We show the combined optimal signal-to-noise ratio $\rho$ from the two independent LISA Michelson outputs. The diamonds correspond to the median of the distribution of $\rho$ over a population of $10^4$ sources, whose parameters have been chosen according to Table~\ref{tab:dataset} (see text for more details). Open diamonds are used for those systems for which we have adopted ``fiducial'' values for the distance and/or chirp mass (the parameters labelled by a superscript star in Table~\ref{tab:dataset}). The error bar is drawn to cover the range between the minimum and maximum value of $\rho$ obtained over the $10^4$ Monte Carlo trials. The right panel refers to the potential verification binaries that have been classified as AM CVn, the left panel to the remaining systems.}
\label{f:plotR}
\end{figure}

\section{Results}

In this section we report the results of the exploration of the SNR and the expected mean square errors on the parameter measurements for the sources in Table~\ref{tab:dataset}. Due to the uncertainty on some of the parameters that determine the actual signal received at the detector -- $\vec{L}$ for all the sources and, in a number of cases, $D$ and $\Mc$ -- we performed Monte Carlo simulations to explore the range of the values that can be obtained during the actual LISA mission. For each binary we consider $10^4$ realisations of the same source with fixed position $\vec{N}$ and GW frequency $f_0$ (according to the values reported in Table~\ref{tab:dataset}). For each realisation, $\vec{L}$ is chosen randomly and drawn from a uniform distribution over $-1 \le \cos\theta_L \le +1 $ and $0 \le \varphi_L \le 2 \pi$. In addition, for those systems for which $D$ is known within a given range (RX J0806.3+1527, V407 Vul, ES Cet and SDSS J1240) the distance is randomly selected from a uniform distribution within the relevant interval. The same is true for $m_1$ for RX J0806.3+1527. For three sources (4U 1820-30, WD 2331+290 and WZ Sge), only limits to $m_1$ and/or $m_2$ are available; in all three cases we fix in the Monte Carlo simulations the value of $m_1$ and $m_2$ to the actual limits reported in Table~\ref{tab:dataset}. Finally, we use for the systems with partially unknown parameters the assumed fiducial values for $D$, $m_1$ and $m_2$, as marked by a superscript star in Table~\ref{tab:dataset}. We assume the standard observation time $T = 1$ yr, and for each of the realisations we compute the optimal SNR, Eq.~(\ref{snr}), and the errors on parameter extraction, Eq.~(\ref{mse}). For the latter, we always assume $f_0$ and $\vec{N}$ to be perfectly known; we therefore compute a four dimensional Fisher information matrix for $A,\,\Phi_0,\,\theta_L$ and $\varphi_L$. We also evaluate the error on the inclination angle $\iota$, cf Eq.~(\ref{e:iota}). In summary for each verification binary we have $10^4$ values of SNR and error on each of the parameters: we then compute the median over those values to obtain a ``typical'' result. For the SNR we also record the minimum and maximum value and use this as a measure of the spread in the results. For $\Delta \vec{\lambda}$ this is not possible because for some choices of $\vec{L}$ some of the parameters become degenerate (and we obtain unphysically large values for $\Delta \vec{\lambda}$). Instead we compute the sample standard deviation of $\Delta \lambda^j$ (for each j = 1,2,3,4) and use it as error bar on the typical values. The results are summarised in Figures 1, 2 and 3.

\begin{figure}
\vspace{0.3cm}
\resizebox{7.5cm}{!}{\includegraphics{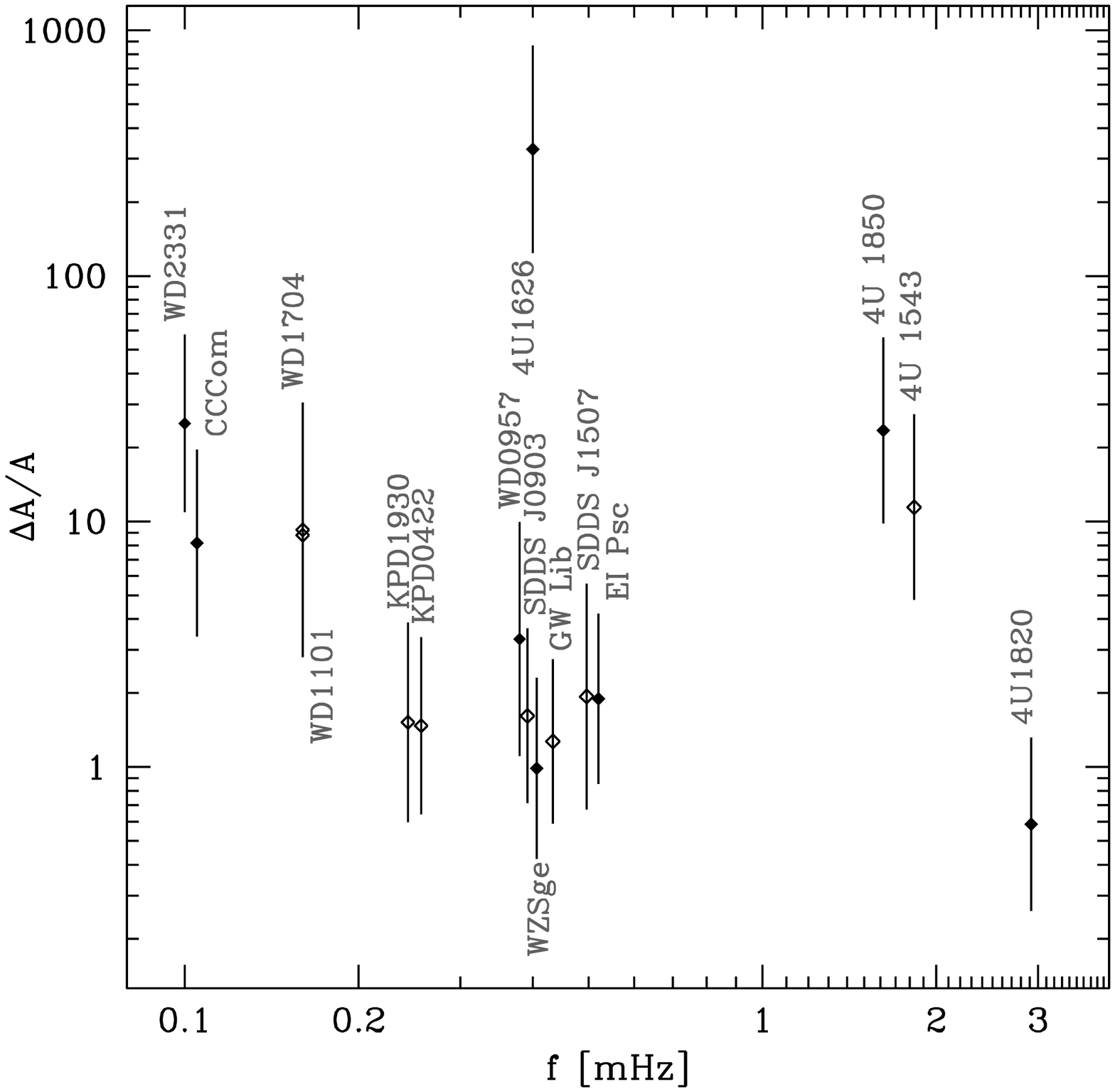}}
\resizebox{7.5cm}{!}{\includegraphics{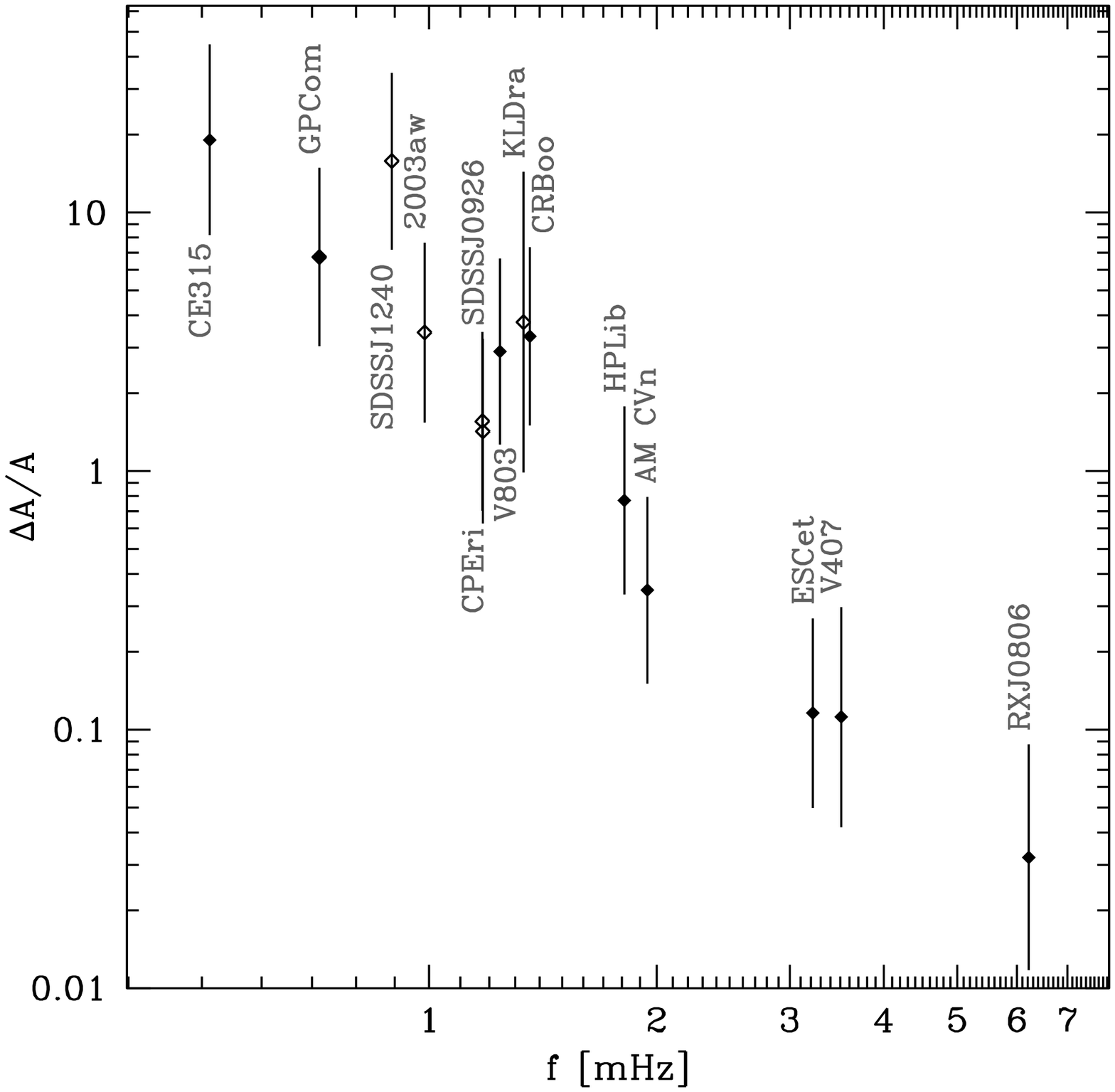}}
\caption{The expected minimum mean square error that characterises the determination of the GW amplitude of the known binaries using one year of LISA data. We plot the relative error $\Delta A/A$ -- the amplitude $A$ is defined according to Eq.~(\ref{e:A}) -- as a function of the GW emission frequency. The diamonds correspond to the median of the distribution of $\Delta A/A$ over a population of $10^4$ sources, whose parameters have been chosen according to Table~\ref{tab:dataset} (see text for more details). Open diamonds are used for those systems for which we have adopted ``fiducial'' values for the distance and/or chirp mass (the parameters labelled by a superscript star in Table~\ref{tab:dataset}). The error bars correspond to the sample standard deviation as measured from the $10^4$ Monte Carlo trials.The right panel refers to the potential verification binaries that have been classified as AM CVn, the left panel to the remaining systems.}
\label{f:plotA}
\end{figure}

We consider first the SNR in order to determine which of the potential verification binaries will be clearly observable. The results are shown in Figure~\ref{f:plotR}. The exact SNR threshold to claim detection depends on a number of details that go beyond the scope of this paper. However, as the search parameters are almost exactly known (source position and gravitational wave frequency), we conservatively set it to $\rho = 5$. If we consider the median value of the SNR, $\bar{\rho}$, as the figure of merit to be compared to the threshold, then only four AM CVn systems are detectable: RXJ0806.3+1527 ($ \bar{\rho} = 62$), V407 Vul ($\bar{\rho} =30$), ES Cet ($\bar{\rho}=19$) and AM CVn ($\bar{\rho}=8$). If the orientation of $\vec{L}$ is favourable, LISA could reach a value of $\rho$ as high as 227, 79, 62 and 13, respectively. It is important to notice that RXJ0806.3+1527 will likely be detected within a week and V407 Vul and ES Cet within a month of science operations. We consider these results robust, because these four systems emit in the frequency band 3 mHz - 7 mHz, in which sources are sufficiently sparse and a possible additional contribution to the noise from confusion as generated by extreme-mass ratio inspirals (if indeed present) degrades the SNR by less than a factor of 2~\cite{Barack2004a}. Furthermore, the use of the Time Delay Interferometry combinations tuned to a given source should increase the SNR by a few tens of percent, see~\cite{Dhur2005} and references therein. We also note that one AM CVn (HP Lib), one LMXB (4U 1820-30), the Cataclysmic Variable WZ Sge and the double degenerate white dwarf KPD 1930+2752 would yield a maximum SNR $\approx 5$, but with the relevant median values below 3. Given the GW emission frequency of these systems, a marginal detection depends on details -- such as degree of signals overlap and actual performance of data analysis algorithms -- that will be known only at the time of the real analysis.

The remaining sources that have been so far included into the potential verification binaries will likely not be observable by LISA; however, several parameters are sufficiently uncertain that a conclusive statement in this respect is premature. We investigated also the likelihood of observing the double radio pulsar PSR J0737-3039: unfortunately, even for a 10 years mission the SNR is only $\approx 0.3$, which suggests that this unique object will be outside LISA's ken.

\begin{figure}
\vspace{0.3cm}
\resizebox{7.5cm}{!}{\includegraphics{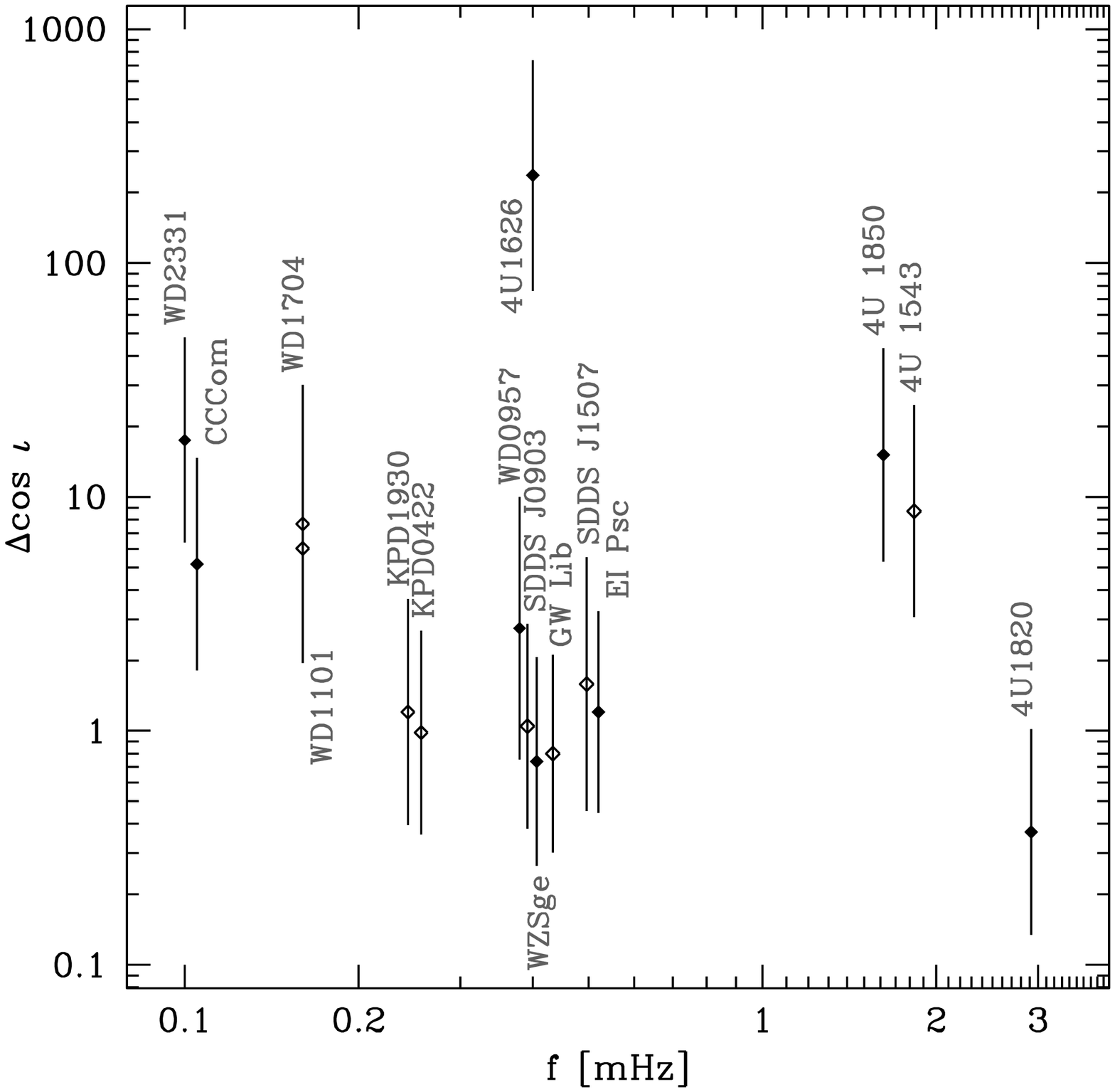}}
\resizebox{7.5cm}{!}{\includegraphics{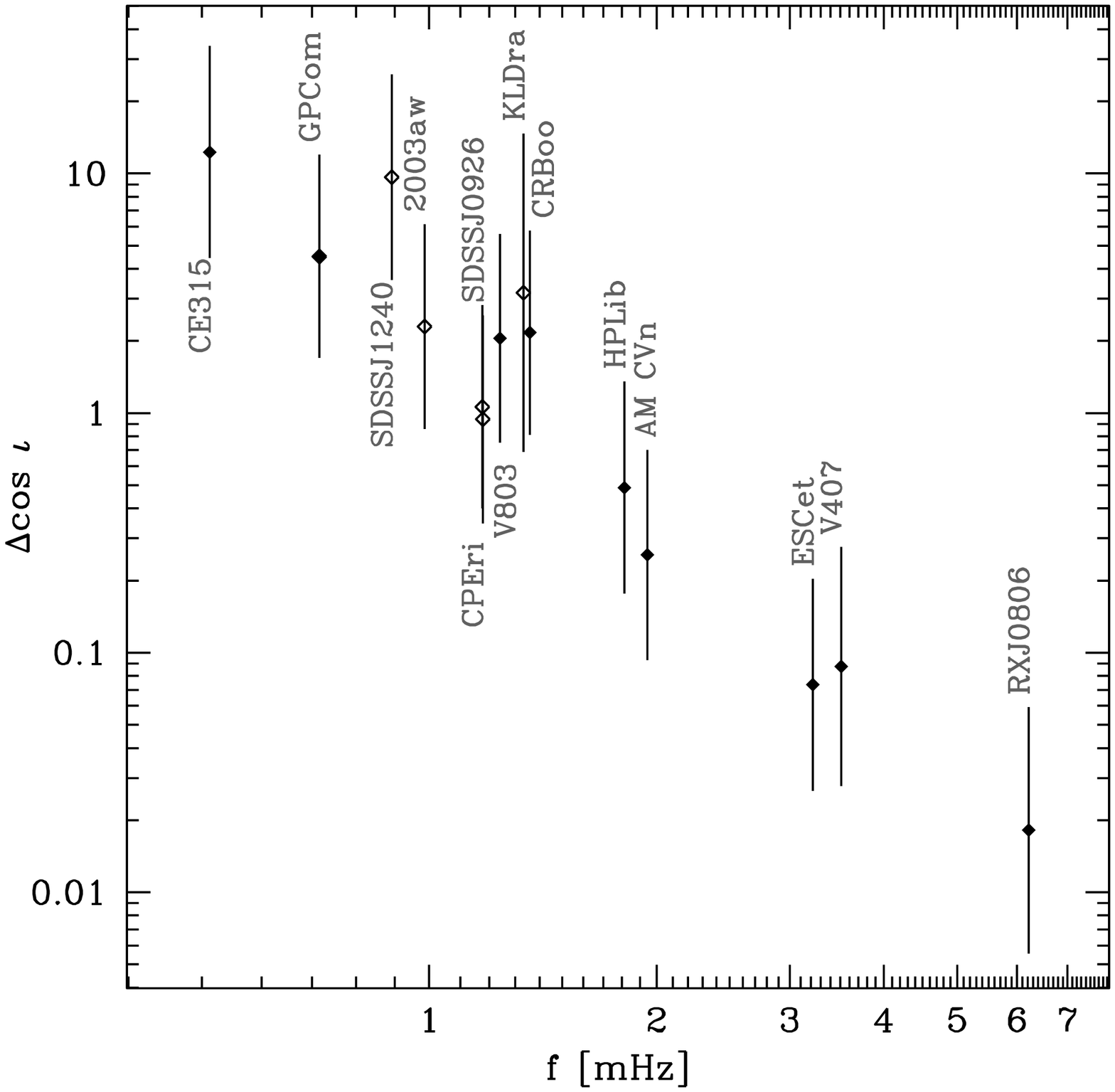}}
\caption{The expected minimum mean square error that characterises the measurement of the inclination angle of the known binaries using one year of LISA data. We plot the error $\Delta \cos\iota$, where $\cos\iota$ is given by Eq.~(\ref{e:iota}), as a function of the GW emission frequency. The symbols are the same as those in Figure~\ref{f:plotA}.}
\label{f:plotI}
\end{figure}

LISA will provide a direct measure of the amplitude $A$, see Eq.~(\ref{e:A}), thus of the ratio $\Mc^{5/3}/D$, and of the elusive (for electromagnetic observations) inclination angle $\iota$, offering a new opportunity to study the physics of compact objects and white dwarfs in particular. Furthermore, if either the chirp mass or the distance are independently known, one could extract information on the other parameter from $A$. The possibility of measuring $\Mc$ is particularly important for AM CVn systems because it could shed new light on the physical effects responsible for the binary orbital evolution~\cite{Stroeer2005a}. In Figures~\ref{f:plotA} and~\ref{f:plotI} we show the range of the expected mean square errors $\Delta A/A$ and $\Delta\cos\iota$. For the clearly detectable verification binaries we have $0.02 \simlt \Delta A/A \simlt 0.3$. Assuming that independent measurements of the distance are of comparable quality by the time LISA is in operation, one would be able to extract information on the chirp mass with a $\approx 10\%$ error. The error on $\cos\iota$ is also sufficiently small, suggesting that one could measure $\cos\iota$ with an error in the range $\approx 0.01 - 0.1$. 

\section{Conclusions}

We have shown that there are indeed four (possibly eight) verification binaries  for LISA. They will be surely detected within the first year of the mission, with one of them, RXJ0806.3+1527, observable by the end of the first week of science operation. High quality observations will be possible for RXJ0806.3+1527, V407 Vul, ES Cet and, to a lesser extent, AM CVn. Our results are clearly affected by uncertainties regarding the actual level of noise during the mission and in particular the effect of astrophysical foregrounds. Furthermore, in our analysis we have not included the full power of the technique of Time Delay Interferometry which will be crucial for the real analysis of the data and will improve the results that we have quoted here by a few tens of percent. Nevertheless, we consider the main conclusions of our paper to be robust. It is now important to develop an end-to-end analysis strategy that can deliver the expected results on real data. This work is currently in progress and will be reported elsewhere.

\section*{Acknowledgements} 

We would like to thank T.~Marsh, G.~Nelemans and G.~Ramsey for several discussions about current observations of galactic binary systems and theoretical models. We would also like to thank J.~Armstrong, M.~Tinto an M.~Vallisneri for sharing some of their preliminary results concerning LISA observations of known binary systems.

\section*{References}

\providecommand{\newblock}{}

\end{document}